\documentclass[journal,twocolumn,10pt,twoside]{IEEEtran}
\DeclareMathAlphabet\mathbfcal{OMS}{cmsy}{b}{n}
\usepackage{cite,amsmath,amssymb,mathrsfs,epsf}
\usepackage{amsfonts}
\usepackage{graphicx}
\usepackage{textcomp}
\usepackage{xcolor}
\usepackage{enumitem}
\usepackage{subfigure}
\usepackage{multirow}
\usepackage{tabularx}
\usepackage{stfloats}
\usepackage{algorithm}
\usepackage{algorithmic}
\usepackage{verbatim}
\usepackage{epsfig}
\usepackage[normalem]{ulem}
\usepackage{float}
\usepackage{hyperref}
\usepackage{color}
\usepackage{mathtools,bm}
\usepackage[outdir=./]{epstopdf}
\usepackage{makecell}
\usepackage{verbatim}
\usepackage{fancyvrb}
\usepackage{fvextra}

\IEEEoverridecommandlockouts

\normalsize

\allowdisplaybreaks

\begin{document}

\title{From Concept to Reality: 5G Positioning with Open-Source Implementation of UL-TDoA in OpenAirInterface}

\author{\IEEEauthorblockN{Adeel Malik\IEEEauthorrefmark{1}, Mohsen Ahadi\IEEEauthorrefmark{2}, Florian Kaltenberger\IEEEauthorrefmark{2}\IEEEauthorrefmark{6}, Klaus Warnke\IEEEauthorrefmark{3}, Nguyen Tien Thinh\IEEEauthorrefmark{2}, \\ Nada Bouknana\IEEEauthorrefmark{2}, Cedric Thienot\IEEEauthorrefmark{1},  Godswill Onche\IEEEauthorrefmark{2}, Sagar Arora\IEEEauthorrefmark{5}} \\
\IEEEauthorblockA{\IEEEauthorrefmark{1}\textit{Firecell}, Nice, France}
\IEEEauthorblockA{\IEEEauthorrefmark{2}\textit{EURECOM}, Sophia Antipolis, France}
\IEEEauthorblockA{\IEEEauthorrefmark{3}\textit{TU Dresden}, Dresden, Germany}
\IEEEauthorblockA{\IEEEauthorrefmark{5}\textit{OpenAirInterface Software Alliance}, Sophia-Antipolis, France}
\IEEEauthorblockA{\IEEEauthorrefmark{6}\textit{Northeastern University}, Boston (MA), USA}

\thanks{The work included in this paper has been supported by the ”France 2030” investment program through the projects 5G-OPERA and GEO-5G.}
}

\maketitle

\newtheorem{axiom}{Axiom}
\newtheorem{lemma}{Lemma}
\newtheorem{corollary}{Corollary}
\newtheorem{theorem}{Theorem}
\newtheorem{prop}{Proposition}
\newtheorem{observation}{Observation}
\newtheorem{definition}{Definition}
\newtheorem{remark}{Remark}

\begin{abstract}

This paper presents, for the first time, an open-source implementation of the 3GPP Uplink Time Difference of Arrival (UL-TDoA) positioning method using the OpenAirInterface (OAI) framework. UL-TDoA is a critical positioning technique in 5G networks, leveraging the time differences of signal arrival at multiple base stations to determine the precise location of User Equipment (UE). This implementation aims to democratize access to advanced positioning technology by integrating UL-TDoA capabilities into both the Radio Access Network (RAN) and Core Network (CN) components of OAI, providing a comprehensive and 3GPP-compliant solution.

The development includes the incorporation of essential protocol procedures, message flows, and interfaces as defined by 3GPP standards. Validation is conducted using two distinct methods: an OAI-RF simulator-based setup for controlled testing and an  O-RAN-based Localization Testbed at EURECOM in real-world conditions. The results demonstrate the viability of this open-source UL-TDoA implementation, enabling precise positioning in various environments. By making this implementation publicly available, the study paves the way for widespread research, development, and innovation in the field of 5G positioning technologies, fostering collaboration and accelerating the advancement of cellular network positioning.


\end{abstract}


\IEEEpeerreviewmaketitle

\section{Introduction}\label{Sec:1}

5G new radio (NR) represents a major leap in wireless communications, significantly improving speed, connectivity, and latency. Beyond enhancing mobile broadband (eMBB), 5G NR enables ultra-reliable low-latency communication (URLLC) and massive machine-type communications (mMTC). A key innovation within the 5G framework is precise positioning and localization, which is anticipated to create new opportunities across sectors such as automotive, healthcare, logistics, and smart city development.


Traditional positioning methods like GPS, Wi-Fi, and Bluetooth Low Energy (BLE) beacons, while commonly used, suffer from limitations regarding accuracy, latency, and reliability. GPS, for instance, provides reliable outdoor coverage but often struggles with accuracy and performance in indoor and densely urban environments, along with high power consumption. Wi-Fi and BLE perform better indoors, but their accuracy can be affected by network density and environmental factors. In contrast, 5G positioning technologies offer superior accuracy, low latency, and consistent performance in both indoor and outdoor scenarios. 5G NR positioning leverages advanced signal characteristics and infrastructure improvements introduced by the new standard. Techniques such as Time Difference of Arrival (TDoA), Angle of Arrival (AoA), and multi-cell Round Trip Time (RTT) are enhanced by 5G's higher frequency bands, massive multiple-input multiple-output (MIMO) technology, and beamforming capabilities. These advancements enable highly accurate and reliable positioning, even in challenging environments like urban canyons and indoor spaces.


Among the various positioning techniques standardized by the 3rd Generation Partnership Project (3GPP)~\cite{3gpp_TS38305}, this paper focuses on the Uplink Time Difference of Arrival (UL-TDoA) method. UL-TDoA determines the location of a User Equipment (UE) by analyzing the time differences in the arrival of uplink signals at multiple base stations. However, the practical implementation of UL-TDoA positioning presents several challenges, including the need for precise synchronization between base stations, advanced signal processing, and robust algorithms to manage multipath propagation and non-line-of-sight conditions. Moreover, proprietary solutions can be costly and rigid, potentially hindering innovation and customization for specific needs.


In this context, open-source software emerges as a vital tool in democratizing access to advanced technologies, promoting innovation, and supporting collaborative development. OpenAirInterface (OAI) is a notable open-source project that offers a comprehensive, 3GPP-compliant implementation of LTE and 5G Radio Access Network (RAN) and Core Network (CN) components. OAI provides a flexible and effective platform for researchers, developers, and network operators to experiment with and deploy advanced communication technologies \cite{Kaltenberger2020}.



\paragraph*{Related work} 
There are several examples where OAI is used as a basis for positioning. In \cite{Li2022}, OAI is combined with a multi-channel SDR and a custom 2x2 patch antenna array for enhanced cell-ID (E-CID) positioning in indoor scenarios, by using RTT and AoA estimates based on the demodulation reference signal (DM-RS).  The HOP-5G project \cite{PeralRosado2024,PeralRosado2024a} sponsored the development of the PRS in OAI for DL-TDoA positioning for FR1 and FR2 and demonstrated the feasibility of this techniques. Based on this work, \cite{Palama2023} built a testbed with three OAI base stations or gNBs and a OAI user equipment exploiting the PRS transmissions for DL-TDoA positioning. More recently, a novel framework is presented in \cite{Mundlamuri2024} to estimate the RTT between a UE and a base station (gNB) with existing 5G NR transmissions.

This paper presents a comprehensive open-source implementation of the 3GPP Uplink Time Difference of Arrival (UL-TDoA) positioning method using the OpenAirInterface (OAI) framework. The primary goal of this work is to integrate UL-TDoA capabilities into both the RAN and CN components of OAI, thereby enabling precise UE positioning within 5G networks. Our implementation enables end-to-end testing of UL-TDoA positioning by incorporating the necessary 3GPP-defined protocol procedures and message exchanges. Key Contributions of this Paper:

\begin{itemize}
    \item Integration of UL-TDoA Support in OAI: We have integrated the UL-TDoA positioning procedure into the OAI framework, enhancing its RAN and CN components to support accurate positioning by implementing the necessary protocol procedures, message flows, and interfaces.
    \item Design and Architecture Insights: We offer detailed insights into integrating UL-TDoA procedures into the OAI architecture, including protocol choices and strategies for 3GPP standards compatibility.
    \item Validation of Implementation: We validate our UL-TDoA implementation using two distinct setups: the OAI-RF simulator-based setup and the O-RAN-based Localization Testbed at EURECOM. These validation methods allow us to demonstrate the functionality and reliability of our implementation under both controlled, simulated conditions and realistic operational scenarios.
    \item Documentation and Tutorial for Implementation: We provide comprehensive documentation and tutorials on how to set up and use the UL-TDoA implementation within the OAI framework. This includes detailed steps for deploying the 5G core and RAN components, configuring the RF simulator, and initiating the positioning procedures.




\end{itemize}

By making this implementation openly available, we aim to provide a valuable resource for the research community and network operators. Our work encourages further exploration and development of UL-TDoA positioning, fostering innovation and collaboration in the field of cellular network positioning technologies.



The following sections will review the technical background of the UL-TDoA positioning framework, highlighting key network entities and protocols. We will then discuss the capabilities of the OpenAirInterface platform and detail the implementation of UL-TDoA positioning within the OAI framework. Next, we will present the validation of our implementation. Finally, we will conclude with a discussion of the implications of our work and future research directions.

\section{3GPP's framework for UL-TDOA based positioning }~\label{section:11}
\begin{figure}
\centering
\includegraphics[width=0.85\linewidth]{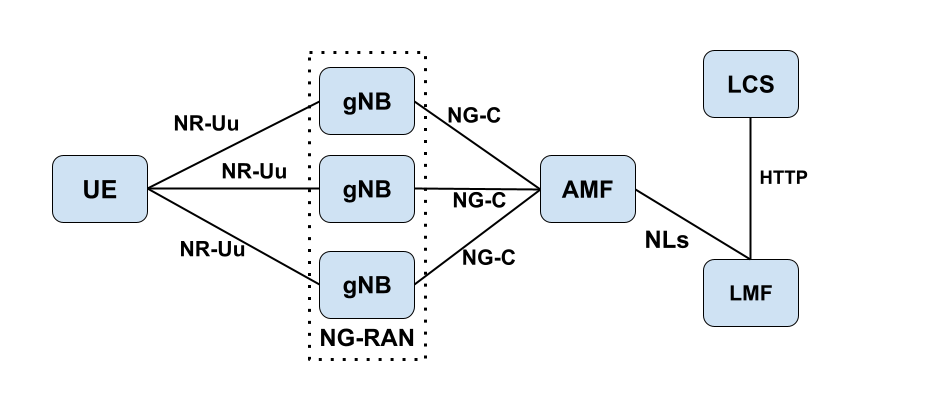}
\caption{UE Positioning Architecture applicable to NG-RAN \cite[Section 5.1]{3gpp_TS38305}}
\label{Pos_Arch}
\end{figure}

The 3GPP framework for positioning in 5G networks involves a combination of network components, interfaces, and positioning methods to provide accurate and reliable location information for User Equipment (UE). The 3GPP standards supporting UL-TDOA positioning are detailed in technical specification TS 38.305\cite{3gpp_TS38305}, which outline the technical requirements for all location services, including UL-TDOA. The framework defined by 3GPP for UL-TDOA-based positioning of a target UE, as well as the delivery of location assistance data to a UE with NG-RAN access in 5G systems, is illustrated in Figure~\ref{Pos_Arch}. The key network components include the UE, Next Generation Radio Access Network (NG-RAN), the Access and Mobility Management Function (AMF), the Location Management Function (LMF), and the Location service LCS entity. These network components are defined as: 
\begin{itemize}
    \item UE: The UE is the mobile device enabled with 5G services (5GS) whose location is being determined.
    \item NG-RAN: The NG-RAN is comprised of at least one Next Generation Node B (gNB). The gNB serves as the base station in 5G networks, facilitating the radio connection between user equipment (UE) and the 5G core network. Each gNB is linked to one or more Transmission Reception Points (TRPs), which are individual antennas or groups of antennas responsible for transmitting and receiving wireless signals.
    \item AMF: The Access and Mobility Management Function (AMF) is an essential part of the 5G core network, tasked with managing access and mobility functions. It oversees UE registration, session management, and ensures seamless mobility across various cells and networks. 
    \item LMF: The Location Management Function (LMF) is a central entity within the 5G core network that oversees location services. It coordinates the collection of TDOA measurements and is responsible for calculating the final position of the UE. 
    \item LCS entity: The Location Services (LCS) entity manages the delivery of location-based services by processing positioning requests from LCS clients. It coordinates with various network elements to ascertain the user equipment's (UE) location. LCS clients can include applications installed on the UE or external systems that require the UE's location data. 
\end{itemize}

The interface defined by the 3GPP that can be used among different network components in Figure~\ref{Pos_Arch} is defined as
\begin{itemize}
    \item NR-Uu: The NR-Uu interface is the radio interface in 5G networks, connecting the User Equipment (UE) to the gNodeB (gNB). It is responsible for the transmission of data, control signals, and user information between the UE and the 5G network. The NR-Uu interface supports various functionalities, including signaling, data transfer, and radio resource management, ensuring efficient communication between the UE and the network. 
    \item NG-C: The NG-C interface in 5G networks connects the gNodeB (gNB) to the Access and Mobility Management Function (AMF). It handles control plane tasks like session and mobility management, ensuring efficient communication between the radio access network and the core network. The NG-C interface is transparent to all UE positioning-related procedures, serving only as a transport link for the NR positioning protocols involved in these processes. In positioning procedures involving the gNB, the NG-C interface transparently transfers positioning requests from the LMF to the gNB and delivers positioning results from the gNB to the LMF.
    \item NLs: The NLs interface connects the Location Management Function (LMF) and the Access and Mobility Management Function (AMF) and is transparent to all UE and gNB positioning procedures. It serves as a transport link for the LTE Positioning Protocol (LPP) and NR Positioning Protocol (NRPPa).
\end{itemize}

\begin{figure}
\centering
\includegraphics[ width=0.85\linewidth]{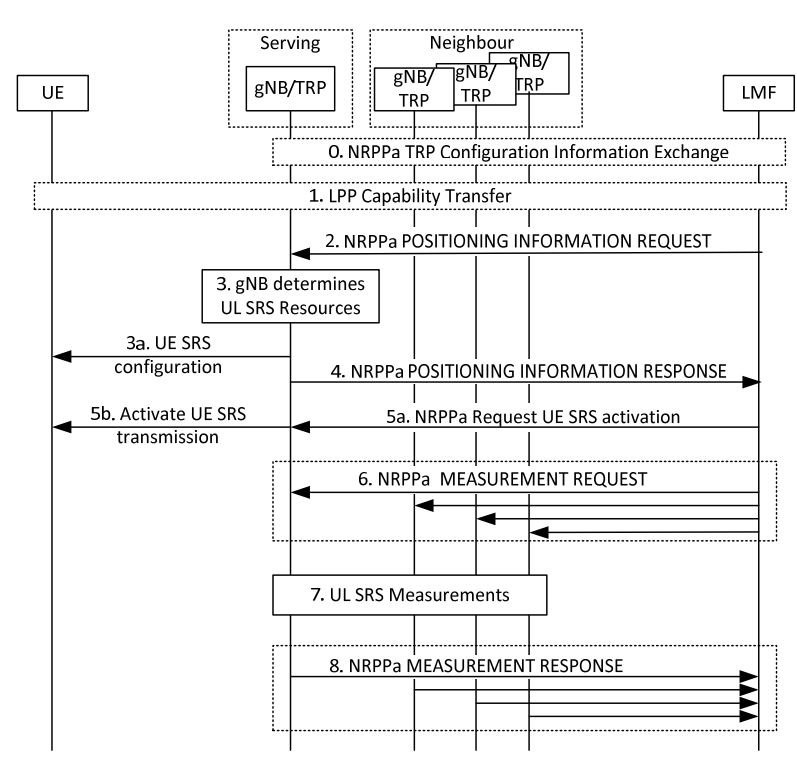} 
\caption{UL-TDoA Positioning Procedure (3GPP) \cite[8.13.3.4]{3gpp_TS38305}.}
\label{fig:ultdoa}
\end{figure}

\subsection{Protocol Required for UL-TDoA Positioning Method in OAI}\label{Sec:model}

The 3GPP specifies various positioning methods in its technical report \cite[Section 8]{3gpp_TS38305}, this paper focuses specifically on the UL-TDoA positioning method and its open-source implementation within the OpenAirInterface framework.

In the UL-TDoA positioning method, the UE location is estimated based on UL-RToA measurements gathered at various gNBs/TRPs for uplink signals from the UE, along with other configuration details. To perform these uplink measurements, participating gNBs/TRPs need to be informed about the characteristics of the SRS signal transmitted by the UE for the required measurement period. These characteristics must remain consistent across the periodic SRS transmissions. The LMF directs the serving gNB to instruct the UE to transmit SRS signals for positioning. The serving gNB then decides on the necessary resources and communicates the SRS configuration to the LMF, which in turn relays this information to the neighboring gNBs/TRPs participating in the UE positioning procedure. 

Figure \ref{fig:ultdoa} defines the UL-TDoA positioning procedure as per 3GPP \cite[Section 8.13]{3gpp_TS38305}. As illustrated in Figure \ref{fig:ultdoa}, the NRPPa protocol is integral to the UL-TDoA positioning procedure, facilitating communication between the LMF and gNB. The NRPPa protocol data units (PDUs) enable the exchange of positioning-related information, including configuration settings, measurement requests, and results. This communication allows the gNBs to precisely measure the TDoA of uplink signals from the UE at various TRPs. The measurements obtained are then utilized by the localization algorithm within the LMF to determine the UE's exact position.

\begin{figure}
\centering
\includegraphics[width=3.5in]{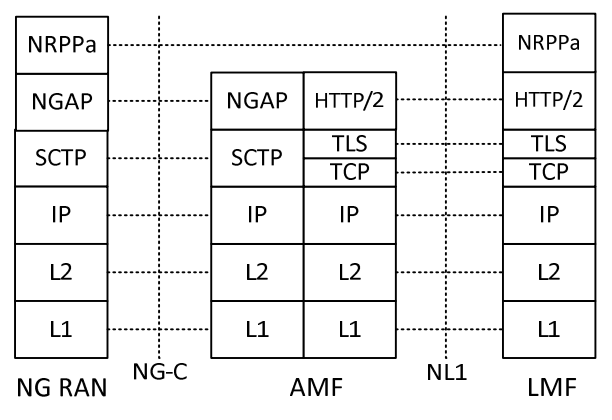}
\caption{Protocol Layering for LMF to NG-RAN Signalling \cite[Section 6.5]{3gpp_TS38305}}
\label{fig:protocol_layer}
\end{figure}

It is important to note that there is no direct connection between the gNB and the LMF; all NRPPa-related messages must pass through the AMF. Figure \ref{fig:protocol_layer} demonstrates the protocol layering required for transferring NRPPa messages between the LMF and gNB. In the next section, we present in detail the messages exchanged by each network element during the NRPPa PDU Transfer process between gNB and LMF.  



\section{Contribution to the OAI and Implementation Status of UL-TDoA}



In this section, we outline the essential components needed within OAI's 5G RAN and 5G Core to facilitate the implementation of the UL-TDoA positioning procedure. A summary of our contributions to OAI's open-source platform is detailed below:
\begin{itemize}
\item Contributions to OAI 5G RAN:
\begin{itemize}
\item Implementation of ToA estimation from SRS
\item Integration of NRPPa functionalities \cite{nrppa}
\item Development of NRPPa PDU Transfer protocol between AMF and gNB \cite{3gpp_TS38413}
\end{itemize}
\item Contributions to OAI 5G Core (LMF):
\begin{itemize}
\item Implementation of LMF-specific procedures \cite{lmf}
\item Integration of NRPPa functionalities \cite{nrppa}
\item Development of NRPPa PDU transfer protocol between AMF and LMF \cite{3gpp_TS29518}
\end{itemize}
Contributions to OAI 5G Core (AMF):
\begin{itemize}
\item Implementation of NRPPa PDU Transfer protocol between AMF and LMF \cite{3gpp_TS29518}
\item Development of NRPPa PDU Transfer protocol between AMF and gNB \cite{3gpp_TS38413}
\end{itemize}
\end{itemize}


These contributions significantly enhance the OAI open-source 5G platform's ability to support precise positioning and location-based services. The following subsection details the specific developments implemented in the OAI codebase to enable the UL-TDoA positioning procedure, which are essential for integrating UL-TDoA functionality within the OAI framework.
\begin{figure*}
\centering
\includegraphics[ width=\textwidth]{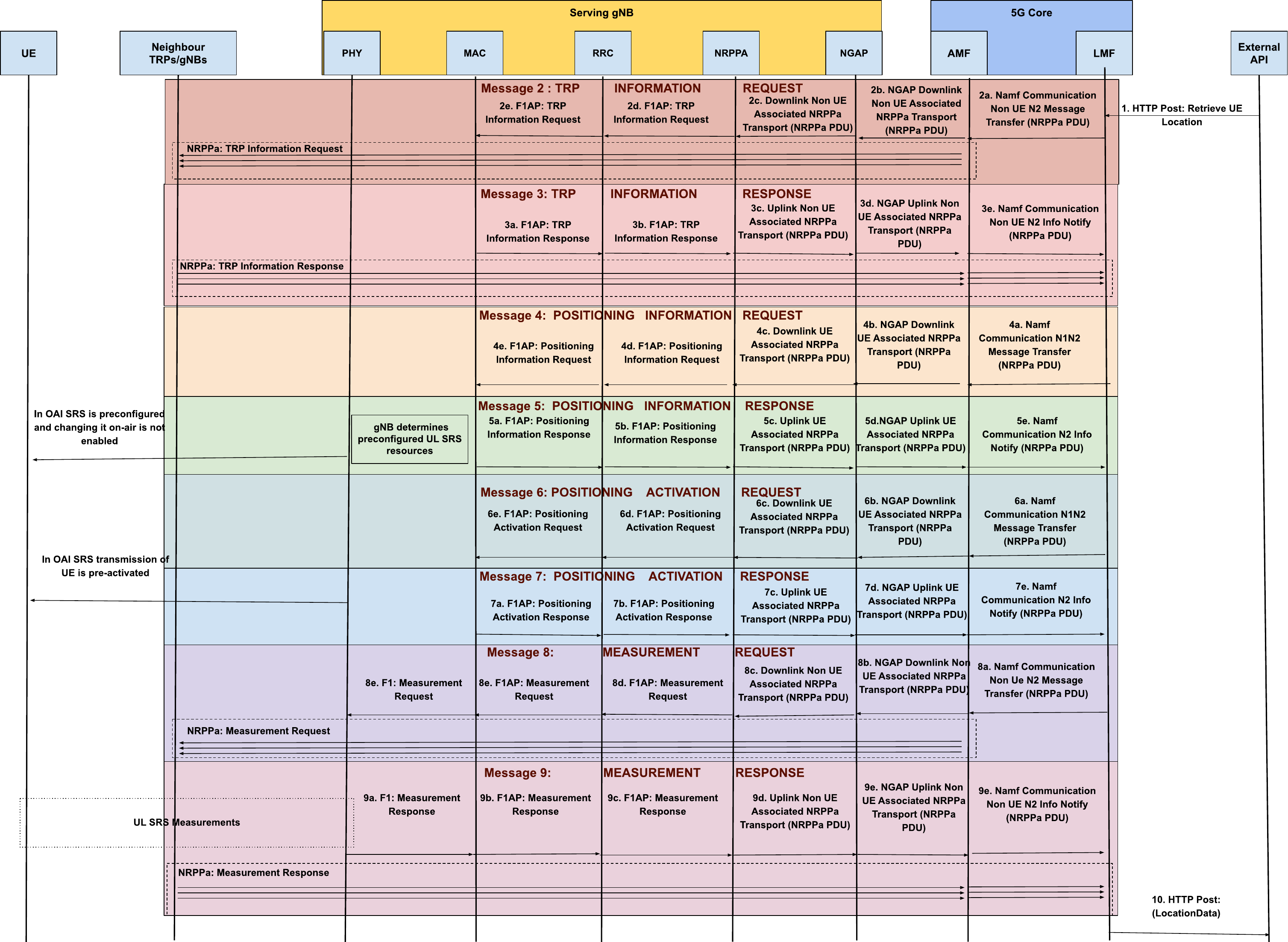}
\caption{End-to-End Implementation of UL-TDoA Positioning Procedure in OAI.}
\label{fig:impl_ultdoa}
\end{figure*}
\subsection{UL-TDoA Positioning Procedure in OAI}


Figure~\ref{fig:impl_ultdoa} provides an overview of our end-to-end implementation of the UL-TDoA positioning procedure in OAI. We will now examine each message exchange in this procedure, highlighting the specific developments we contributed to the OAI framework.

\noindent \textit{\textbf{Message 1 Initiation of Location Request:}} 

\begin{figure}
\centering
\includegraphics[width=3.5in]{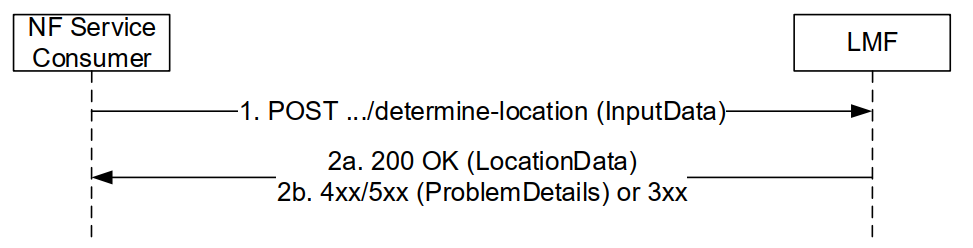}
\caption{Determine Location Request \cite[Figure 5.2.2.2.2-1]{lmf}}
\label{fig:determineLoc}
\end{figure}

The UL-TDoA positioning procedure begins with an external API generating a location information request for a target UE for the LMF. As shown in Figure ~\ref{fig:determineLoc} to initiate the location procedure, the external API (which can be an advanced API or a simple one-line command) sends the HTTP post request to \textit{determine-location} API of LMF, where the request contains a data structure of type \textit{InputData} \cite[Section 6.1.6.2.2]{lmf}. To enable the exchange of this message, we first developed the 3gpp-compliant LMF framework and integrated it into OAI's core network framework\cite{oai_lmf}. Then within that LMF framework, we develop the determine-location API \cite[Section 6]{lmf} to handle the positioning request following the 3gpp standard. 

The 3GPP outlines multiple identification methods \cite[Section 6.1.6.2.2]{lmf} for initiating the positioning of a target UE. In our implementation, we utilized the \textit{Subscription Permanent Identifier} (SUPI) to identify the target UE. The SUPI functions as a permanent identifier for subscribers within the network, supporting various authentication, authorization, and identification tasks in 5G networks. Specifically, in our implementation, the SUPI is represented by the International Mobile Subscriber Identity (IMSI), a commonly used identifier in mobile networks. The key information elements included in the InputData to initiate the location request for our implementation are as follows:



\begin{itemize}
    \item SUPI
    \item NR Cell Global Identifier (NCGI)
    \item Periodic Event Info
\end{itemize}
Once the LMF receives the determine-location request for a specific UE it initiates the procedure to retrieve all the required information from gNBs for the UL-TDOA positioning method. The procedure consists of several NRPPA messages (Messages 2-9 of Figure~\ref{fig:impl_ultdoa}) exchanged between LMF and gNB(s). 


\noindent \textit{\textbf{Message 2 TRP Information Request:}}

The LMF initiates the UL-TDOA positioning method by sending a TRP information request \cite[Section 9.1.1.14]{nrppa} to obtain the necessary TRP information for UE positioning. The key information elements included in our implementation of the TRP information request, based on the list provided by 3GPP \cite[Section 9.1.1.14]{nrppa}, are as follows:


%
\begin{itemize}
    \item Message Type	
    \item NRPPa Transection ID
\end{itemize}

The TRP information request is part of the non-UE-associated NRPPa procedures. In our implementation, the AMF forwards this non-UE-associated message to all the gNBs connected to it. The TRP information request message terminates in the MAC handler of the gNB. We have added processing for this message in the LMF, AMF, gNB-NGAP, gNB-NRPPA, gNB-RRC, and gNB-MAC of OAI's open-source 5G framework.


\begin{enumerate}[label=\textbf{Message 2.\alph*.},align=left] 
\item The LMF creates an NRPPa PDU for the TRP Information Request and forwards it to the AMF by initiating the \textit{Namf Communication Non-UE N2 Message Transfer} procedure, carrying the NRPPa PDU. 
\item The AMF then initiates the \textit{NGAP Downlink Non-UE Associated NRPPa Transport} procedure, carrying the NRPPa PDU, and sends it to all the gNBs connected to the AMF. 
\item Upon receiving the message from the AMF, each gNB processes the NGAP message and forwards the NRPPa PDU to the \textit{Downlink Non-UE Associated NRPPa Transport} handler in the gNB-NRPPa thread. 
\item The gNB-NRPPa thread decodes the NRPPa PDU and forwards an \textit{F1AP: TRP Information Request}\footnote{The F1 Application Protocol (F1AP) is used to manage and control the interaction between the gNB Central Unit (gNB-CU) and the gNB Distributed Unit (gNB-DU) in 5G NR architecture when the gNB is operating in split mode. We have added the basic framework for all positioning messages 2-9 of Figure~\ref{fig:impl_ultdoa} in OAI's framework to support split mode; however, the complete processing of these messages has only been implemented for monolithic mode. Our goal is to extend our work to support the positioning procedure in split mode.} message to the gNB-RRC. 
\item The gNB-RRC thread processes the request and forwards the \textit{F1AP: TRP Information Request} to the gNB-MAC. 
\end{enumerate}

\noindent \textit{\textbf{Message 3 TRP Information Response:}}

All gNBs participating in the positioning procedure prepare a TRP information response and send it to the LMF, providing essential data such as configuration and location, which are crucial for accurate UE positioning. The key information elements included in our implementation of the TRP information response, based on the list provided by 3GPP \cite[Section 9.1.1.15]{nrppa}, are as follows:

\begin{itemize}	
\item Message Type
\item NRPPa Transection ID
\item TRP Information
\begin{itemize}
\item TRP ID
\item NG-RAN CGI
\begin{itemize}
\item PLMN Identity
\end{itemize}
\item Geographical Coordinates
\begin{itemize}
\item Relative Cartesian Location
\item Location Uncertainty
\end{itemize}
\end{itemize}
\end{itemize}


The TRP Information Response message terminates at the LMF, and we have implemented processing for this message in the LMF, AMF, gNB-NGAP, gNB-NRPPA, gNB-RRC, and gNB-MAC components of OAI's open-source 5G framework.

\begin{enumerate}[label=\textbf{Message 3.\alph*.},align=left] 
\item The gNB-MAC handler processes the TRP Information Request, creates an \textit{F1AP: TRP Information Response} containing the relevant TRP information, and forwards it to the gNB-RRC for further processing. 
\item The gNB-RRC handler processes the \textit{F1AP: TRP Information Response} received from the gNB-MAC and forwards it to the gNB-NRPPa handler. 
\item The gNB-NRPPa handler processes the \textit{F1AP: TRP Information Response}, extracts key information, creates an NRPPa PDU, and sends it to the gNB-NGAP handler by initiating the \textit{Uplink Non-UE Associated NRPPa Transport} procedure. 
\item The gNB-NGAP then initiates the \textit{NGAP Uplink Non-UE Associated NRPPa Transport} procedure and forwards the NRPPa PDU to the AMF. 
\item The AMF forwards the NRPPa PDU to the LMF by initiating the \textit{Namf Communication Non-UE N2 Info Notify} procedure. 
\end{enumerate}

\noindent \textit{\textbf{Message 4 Positioning Information Request:}}
After receiving the TRP Information Responses from all gNBs involved in the positioning process, the LMF initiates the retrieval of UL-SRS configuration details for the target UE. It then sends a Positioning Information Request message to the serving gNB. In response, the serving gNB allocates the necessary resources for UL-SRS and configures the target device with the appropriate UL-SRS resource sets. The key information elements included in our implementation of the Positioning Information Request, as outlined by 3GPP \cite[Section 9.1.1.10]{nrppa}, are as follows:


\begin{itemize}	
\item Message Type
\item NRPPa Transaction ID
\item Requested SRS Transmission Characteristics
\end{itemize}



The Positioning Information Request is part of the UE-associated NRPPa procedures. Therefore, the AMF will forward this message only to the gNB that is serving the UE. The Positioning information request message terminates in the MAC handler of the gNB. We have added processing for this message in the LMF, AMF, gNB-NGAP, gNB-NRPPA, gNB-RRC, and gNB-MAC of OAI’s open-source 5G framework.

\begin{enumerate}[label=\textbf{Message 4.\alph*.},align=left] 
\item The LMF creates an NRPPa PDU for the Positioning Information Request and forwards it to the AMF by initiating the \textit{Namf Communication N1N2 Message Transfer} procedure, carrying the NRPPa PDU. 
\item The AMF then initiates the \textit{NGAP Downlink UE Associated NRPPa Transport} procedure, carrying the NRPPa PDU, and sends it to the serving gNB. 
\item Upon receiving the message from the AMF, the gNB processes the NGAP message and forwards the NRPPa PDU to the \textit{Downlink UE Associated NRPPa Transport} handler in the gNB-NRPPa thread. 
\item The gNB-NRPPa thread decodes the NRPPa PDU and forwards a \textit{F1AP: Positioning Information Request} message to the gNB-RRC. 
\item The gNB-RRC thread processes the request and forwards the \textit{F1AP: Positioning Information Request} to the gNB-MAC. 
\end{enumerate}

\noindent \textit{\textbf{Message 5 Positioning Information Response:}}
After processing the Positioning Information Request, the serving gNB prepares a Positioning Information Response that includes the UL SRS configuration for the target UE. In the current OAI RAN framework, the SRS configuration is predefined in the gNB settings and does not allow for dynamic adjustments for individual UEs. Consequently, when the serving gNB receives the Positioning Information Request, it disregards the specific configuration requested by the LMF and instead returns the existing UL SRS configuration that has already been applied to the target UE.  The key information elements included in our implementation of the Positioning Information response, as outlined by 3GPP \cite[Section 9.1.1.11]{nrppa}, are as follows:

\begin{itemize}	
\item  Message Type
\item  NRPPa Transaction ID
\item  SRS Configuration
\end{itemize}



The Positioning Information Response message terminates at the LMF, and we have implemented processing for this message in the LMF, AMF, gNB-NGAP, gNB-NRPPA, gNB-RRC, and gNB-MAC components of OAI's open-source 5G framework.

\begin{enumerate}[label=\textbf{Message 5.\alph*.},align=left] 
\item The gNB-MAC handler processes the Positioning Information Request, creates an \textit{F1AP: Positioning Information Response} containing the SRS configuration, and forwards it to the gNB-RRC for further processing. 
\item The gNB-RRC handler processes the \textit{F1AP: Positioning Information Response} received from the gNB-MAC and forwards it to the gNB-NRPPa handler. 
\item The gNB-NRPPa handler processes the \textit{F1AP: Positioning Information Response}, extracts key information, creates an NRPPa PDU, and sends it to the gNB-NGAP handler by initiating the \textit{Uplink UE Associated NRPPa Transport} procedure. 
\item The gNB-NGAP then initiates the \textit{NGAP Uplink UE Associated NRPPa Transport} procedure and forwards the NRPPa PDU to the AMF. 
\item The AMF forwards the NRPPa PDU to the LMF by initiating the \textit{Namf Communication N2 Info Notify} procedure. 
\end{enumerate}

\noindent \textit{\textbf{Message 6 Positioning Activation Request:}}

Once the LMF is informed of the SRS configuration for the target UE, it initiates the activation of UE SRS transmission by sending an NRPPa Positioning Activation Request message to the serving gNB. The key information elements included in our implementation of the Positioning Activation Request, as outlined by 3GPP \cite[Section 9.1.1.12]{nrppa}, are as follows:

\begin{itemize}	
\item Message Type
\item NRPPa Transection ID
\item SRS Resource Set ID
\item SRS Resource Trigger
\end{itemize}





The Positioning Activation Request is part of the UE-associated NRPPa procedures. Therefore, the AMF will forward this message only to the gNB that is currently serving the UE. The Positioning Activation request message terminates in the MAC handler of the gNB. We have added processing for this message in the LMF, AMF, gNB-NGAP, gNB-NRPPA, gNB-RRC, and gNB-MAC of OAI’s open-source 5G framework.


\begin{enumerate}[label=\textbf{Message 6.\alph*.},align=left] 
\item The LMF creates an NRPPa PDU for the Positioning Activation Request and forwards it to the AMF by initiating the \textit{Namf Communication N1N2 Message Transfer} procedure, carrying the NRPPa PDU. 
\item The AMF then initiates the \textit{NGAP: Downlink UE-associated NRPPa Transport} procedure, carrying the NRPPa PDU, and sends it to the serving gNB.
\item Upon receiving the message from the AMF, the gNB processes the NGAP message and forwards the NRPPa PDU to the \textit{Downlink UE-associated NRPPa Transport} handler in the gNB-NRPPa thread. 
\item The gNB-NRPPa thread decodes the NRPPa PDU and forwards a \textit{F1AP: Positioning Activation Request} message to the gNB-RRC. 
\item The gNB-RRC thread processes the request and forwards the \textit{F1AP: Positioning Activation Request} to the gNB-MAC.
\end{enumerate}

\noindent \textit{\textbf{Message 7 Positioning Activation Response:}}
 In the current OAI RAN framework, UE SRS transmission is preactivated, and real-time modifications to a UE's SRS transmission are not supported. As a result, when the gNB receives the request, it generates a Positioning Activation Response and sends it back to the LMF, indicating that the target device is transmitting the SRS. The key information elements included in our implementation of the Positioning Activation Response, as outlined by 3GPP \cite[Section 9.1.1.18]{nrppa}, are as follows:

\begin{itemize}	
\item Message Type
\item NRPPa Transaction ID
\item Criticality Diagnostics
\item System Frame Number
\item Slot Number
\end{itemize}




The Positioning Activation Response message terminates at the LMF, and we have implemented processing for this message in the LMF, AMF, gNB-NGAP, gNB-NRPPA, gNB-RRC, and gNB-MAC components of OAI's open-source 5G framework. Although UE SRS transmission is preactivated, allowing the Positioning Activation Request to be handled entirely at the gNB-NRPPa with a response generated there, we have included the processing of this message in both the gNB-RRC and gNB-MAC for completeness. This approach ensures future compatibility with potential OAI code developments that may support on-the-fly SRS activation.



\begin{enumerate}[label=\textbf{Message 7.\alph*.},align=left] 
\item The gNB-MAC handler processes the Positioning Activation Request, creates an \textit{F1AP: Positioning Activation Response}, and forwards it to the gNB-RRC for further processing. 
\item The gNB-RRC handler processes the \textit{F1AP: Positioning Activation Response} received from the gNB-MAC and forwards it to the gNB-NRPPa handler. 
\item The gNB-NRPPa handler processes the \textit{F1AP: Positioning Activation Response}, extracts key information, creates an NRPPa PDU, and sends it to the gNB-NGAP handler by initiating the \textit{Uplink UE-associated NRPPa Transport} procedure. 
\item The gNB-NGAP initiates the \textit{NGAP: Uplink UE-associated NRPPa Transport} procedure and forwards the NRPPa PDU to the AMF. 
\item The AMF forwards the NRPPa PDU to the LMF by initiating the \textit{Namf Communication N2 Information Notify} procedure.
\end{enumerate}

\noindent \textit{\textbf{Message 8 Measurement Request:}}



Once the Positioning Activation Response is received by the LMF, it initiates the retrieval of ToA measurements from all the gNBs/TRPs involved in the positioning of the target UE. The LMF sends a Measurement Request to each participating gNB, which includes the SRS configuration previously received from the serving gNB in the Positioning Information Response for the target UE. The participating gNBs then use this configuration to perform the ToA measurements and send the results back to the LMF. The key information elements included in our implementation of the measurement Request, as outlined by 3GPP \cite[Section 9.1.4.1]{nrppa}, are as follows:

\begin{itemize}	
\item Message Type
\item NRPPa Transection ID
\item ITRP ID
\item SRS Configuration
\end{itemize}


The measurement request is part of the non-UE-associated NRPPa procedures. In our implementation, the AMF forwards this non-UE-associated message to all the gNBs connected to it\footnote{In future work, we plan to enable the LMF and AMF to send Measurement Requests to specific gNBs. This approach allows for the selection of a targeted subset of gNBs to participate in the positioning process and provide measurements, optimizing resource usage and potentially improving accuracy.}. The measurement request message terminates in the PHY handler of the gNB. We have added processing for this message in the LMF, AMF, gNB-NGAP, gNB-NRPPA, gNB-RRC, gNB-MAC, and gNB-PHY of OAI's open-source 5G framework.



\begin{enumerate}[label=\textbf{Message 8.\alph*.},align=left] 
\item The LMF creates an NRPPa PDU for the Measurement Request and forwards it to the AMF by initiating the \textit{Namf Communication Non-UE N2 Message Transfer} procedure, which carries the NRPPa PDU. 
\item The AMF then initiates the \textit{NGAP Downlink Non-UE Associated NRPPa Transport} procedure, carrying the NRPPa PDU, and sends it to all the gNBs connected to the AMF. 
\item Upon receiving the message from the AMF, each gNB processes the NGAP message and forwards the NRPPa PDU to the \textit{Downlink Non-UE Associated NRPPa Transport} handler in the gNB-NRPPa thread. 
\item The gNB-NRPPa thread decodes the NRPPa PDU and forwards an \textit{F1AP: Measurement Request} message to the gNB-RRC. 
\item The gNB-RRC thread processes the request and forwards the \textit{F1AP: Measurement Request} to the gNB-MAC. 
\item The gNB-MAC thread processes the request and forwards the \textit{FAPI: Measurement Request} to the gNB-PHY.
\end{enumerate}

\noindent \textit{\textbf{Message 9 Measurement Response:}}


All gNBs that have received the Measurement Request and are involved in the positioning procedure generate a Measurement Response and send it to the LMF, providing crucial ToA measurements necessary for accurate UE positioning. The precision and resolution of these ToA measurements are essential for achieving high positioning accuracy, as any errors in the ToA measurements can significantly affect the overall accuracy. To enhance measurement precision, we have integrated a state-of-the-art channel estimation and ToA estimation procedure into the gNB. A detailed overview of this integration is provided in Section \ref{sec:toa_estimation}. The key information elements included in our implementation of the measurement response, based on the list provided by 3GPP \cite[Section 9.1.4.2]{nrppa}, are as follows:

\begin{itemize}	
\item Message Type
\item NRPPa Transection ID
\item TRP ID
\item Measurement Result
\begin{itemize}
\item UL RTOA Measurement
\item gNB Rx-Tx Time Difference
\end{itemize}
\end{itemize}

%



The measurement Response message terminates at the LMF, and we have implemented processing for this message in the LMF, AMF, gNB-NGAP, gNB-NRPPA, gNB-RRC,
gNB-MAC, and gNB-PHY components of OAI’s open-source 5G framework.

\begin{enumerate}[label=\textbf{Message 9.\alph*.},align=left] 
\item The gNB-PHY handler processes the  Request, creates an \textit{FAPI: Measurement Response} containing the ToA measuement and corresponding TRP information, and forwards it to the gNB-MAC for further processing.
\item The gNB-MAC handler processes the  Response, creates an \textit{F1AP: Measurement Response} containing the relevant TRP information, and forwards it to the gNB-RRC for further processing. 
\item The gNB-RRC handler processes the \textit{F1AP: Measurement Response} received from the gNB-MAC and forwards it to the gNB-NRPPa handler. 
\item The gNB-NRPPa handler processes the \textit{F1AP: Measurement Response}, extracts key information, creates an NRPPa PDU, and sends it to the gNB-NGAP handler by initiating the \textit{Uplink Non-UE Associated NRPPa Transport} procedure. 
\item The gNB-NGAP then initiates the \textit{NGAP Uplink Non-UE Associated NRPPa Transport} procedure and forwards the NRPPa PDU to the AMF. 
\item The AMF forwards the NRPPa PDU to the LMF by initiating the \textit{Namf Communication Non-UE N2 Info Notify} procedure. 
\end{enumerate}

\noindent \textit{\textbf{Message 10 Location Response:}}

The Location Management Function (LMF) receives Time of Arrival (ToA) measurements (uLRTOAmeas) from each gNB and TRP, along with the TRP information response containing the relative Cartesian coordinates (x, y, z) of each TRP, which are essential for computing the user's position. Before using these ToA measurements, the LMF must map the reported uLRTOAmeas from the gNB to actual measured values using predefined tables from ETSI TS 138 133. For each location determination request, the LMF converts the array of ToA values into Time Difference of Arrival (TDoA) values, referencing the TRP with the strongest Received Signal Received Power (RSRP). Using the known TRP locations, a localization algorithm within the LMF—currently utilizing linear and nonlinear least squares solutions—calculates the user's location and returns it to the determine-location API. This localization function is designed to accept TDoA values and TRP positions as input, producing an estimated user position as output, allowing for the integration of any localization algorithm that follows this input-output format into the LMF.


Once the LMF has calculated the position of the target UE, it initiates the process to provide a response to the External API that initiated the positioning request, as illustrated in Figure \ref{fig:determineLoc}. This response includes a data structure of type \textit{LocationData} \cite[Section 6.1.6.2.3]{lmf}. The key information elements included in our implementation of the positioning response (i.e., LocationData), based on the list provided by 3GPP \cite[Section 6.1.6.2.3]{lmf}, are as follows:
\begin{itemize}
    \item Geographical coordinates
    \item Relative Cartesian Location
\end{itemize}


This concludes our explanation of the end-to-end implementation of the UL-TDoA positioning procedure in OAI, as illustrated in Figure~\ref{fig:impl_ultdoa}. In the following section, we introduce our state-of-the-art ToA estimation procedure, which has been integrated into OAI's gNB.

\subsection{Enabling ToA estimation in OAI's gNB}\label{sec:toa_estimation}

\subsubsection{gNB-PHY}
The PHY layer involves two primary processes: channel estimation and subsequent Time of Arrival (ToA) estimation based on the pilot symbols of the Sounding Reference Signal (SRS). The SRS is a wide-band reference signal transmitted by the UE in the uplink. The SRS for positioning is generated using the Zadoff-Chu sequence \cite{3gpp_ts_38_211}, similar to the SRS for communication, although they can be configured differently.
During the SRS channel estimation, essential parameters such as the OFDM symbol size $N_{\text{OFDM}}$, subcarrier offset $k$, number of antennas $N_{\text{RX}}$, and antenna port configurations $P_{\text{RX}}$ are extracted from the gNB structure. For each receive antenna $n \in \{1, \dots, N_{\text{RX}}\}$ and antenna port $p \in \{1, \dots, P_{\text{RX}}\}$, Least Squares (LS) estimation is performed by correlating the received SRS signal $\mathbf{Y}_{SRS_{n,p}}$ with the generated signal $\mathbf{X}_{SRS_{n,p}}$ to estimate the channel $\hat{\mathbf{H}}_{{n,p,l}}$:
\begin{equation}
    \hat{\mathbf{H}}_{n,p,l}[k] =  \frac{\mathbf{Y}_{SRS_{n,p,l}}[k]}{\mathbf{X}_{SRS_{n,p,l}}[k]}
\end{equation}
Since the SRS can be mapped to multiple consecutive OFDM symbols $l \in \{1, \dots, N_{symb}^{SRS}\}$, during the channel estimation process, the SRS channel is estimated over all the symbols.\\
The channel estimate is then interpolated to refine the estimates obtained from the LS estimation process using filter vectors $\mathbf{F}_{\text{start}}$, $\mathbf{F}_{\text{middle}}$, or $\mathbf{F}_{\text{end}}$, depending on the position within the OFDM symbol, as follows:
\begin{equation}
   \hat{\mathbf{H}}_{n,p,l}^{\text{interp}}[k]\! \!= \!\!
    \begin{cases} 
        \!\mathbf{F}_{\text{start}} \cdot \hat{\mathbf{H}}_{n,p,l}[k] & \!\!\!\!\text{if } \!k = 0 \text{ or } k < K_{\text{TC}} \\
       \! \mathbf{F}_{\text{middle}} \cdot \hat{\mathbf{H}}_{n,p,l}[k] & \!\! \!\!\text{if } \!k \neq 0 \text{ and } (k + K_{\text{TC}}) < N_{\text{sc}} \\
       \! \mathbf{F}_{\text{end}} \cdot \hat{\mathbf{H}}_{n,p,l}[k] & \!\!\!\!\text{if } \!(k + K_{\text{TC}}) \!\geq\! N_{\text{sc}} \!\text{ or }\! k \!= \!M_{\text{sc}}\! -\! 1
    \end{cases}
\end{equation}
where $K_{\text{TC}}$ is the comb size parameter, $M_{\text{sc}}$ is the number of subcarriers in an SRS sequence, and $N_{\text{sc}}$ is the total number of subcarriers in an OFDM symbol.\\
Next, frequency domain channel oversampling and conversion of the estimated channel from the frequency domain to the time domain are performed using an Inverse Fast Fourier Transform (IFFT):
\begin{gather}
\hat{\mathbf{H}}_{n,p,l}^{\text{oversamp}} =
\begin{bmatrix}
\hat{\mathbf{H}}_{n,p,l}^{\text{interp}}(1 : N_{\text{fft}}/2),\\
\mathbf{zeros}(1, L \times N_{\text{fft}} - N_{\text{fft}}),\\
\hat{\mathbf{H}}_{n,p,l}^{\text{interp}}(N_{\text{fft}}/2 + 1: \text{end})
\end{bmatrix}
\end{gather}
where $L$ is the oversampling factor, and $N_{\text{fft}}$ is the IFFT length. The IFFT operation is defined as:
\begin{equation}
    \hat{\mathbf{h}}^{\text{oversamp}}_{n,p,l}(t) = \frac{1}{L\times N_{\text{fft}}} \sum_{k=0}^{L \times N_{\text{fft}}-1} \hat{\mathbf{H}}_{n,p,l}^{\text{oversamp}}[k] e^{i 2 \pi k t / (L \times N_{\text{fft}})}
\end{equation}
where $t$ denotes the time-domain index.
The ToA is estimated by identifying the index $\tau_{n}^{\text{peak}}$ corresponding to the maximum of the magnitude squared of the channel impulse response averaged over all the SRS OFDM symbols:
\begin{equation}
   \tau_{n}^{\text{peak}} = \arg \max_{t} \{\frac{1}{N_{\text{symb}}^{\text{SRS}}}\sum_{l=1}^{N_{symb}^{SRS}}\sum_{p=1}^{P_{RX}} \left|\hat{h}_{n,p,l}(t)\right|^2 \}
\end{equation}

The ToA in seconds can be calculated as:

\begin{equation}
    \text{ToA}_{n} = \left(\frac{\tau_{n}^{\text{peak}}}{L}\right) \cdot T_s
\end{equation}

Where $T_s$ is the sampling period.
\paragraph{FAPI interface}
The FAPI interface is a standardized interface between the PHY and the MAC specified by the small cell forum \cite{SCF2021}. The SRS measurements are sent from PHY to the MAC via the SRS.indication message. This message includes a field called ``Timing advance offset in nanoseconds,'' which is meant to be used for UL TDoA positioning measurements. However, the standard does not foresee that there can be multiple such measurements from multiple TRPs connected to the same DU. We solved this issue by adding a new value for SRS type (5) and a new report type (``Localization'') to the SRS indication message. These additions are allowed in the current specification and do not break the compatibility with implementations that do not support this report type. The actual values of the Timing advance offsets in ns from the multiple TRPs are sent in the report-tagged list value (TLV) directly as an array in the value field, where each value uses 16bit.  

\subsubsection{gNB-MAC}

The MAC is responsible for scheduling the SRS and for handling the SRS indication messages from the PHY.

The scheduling is different depending if the UE is connected to the same MAC (serving cell measurements) or to a different MAC (neighbor cell measurements). 

The serving cell measurements are straightforward as the SRS scheduling is already activated at the PDU session establishment during the connection procedure. Upon reception of an F1 positioning measurement request, the MAC simply takes the latest ToA measurement it received from the PHY and generates the F1 positioning measurement response.

In the case of a neighbor cell measurement, the MAC first has to activate the SRS measurements. For this kind of measurement we have also had to slightly modify the usage of the FAPI interface as the current interface does not support this usage. We have introduced a special RNTI reserved for this usage. The RNTI is used in the SRS PDU of the UL\_TTI request sent from the MAC to the PHY and also in the SRS indication sent back from the PHY to the MAC. This way we can associate the measurement with the measurement request message. 

\section{Validation of our Open-Source UL-TDoA positioning}
In this section, we validate our open-source UL-TDoA positioning implementation within the OpenAirInterface (OAI) framework. We conducted end-to-end protocol testing and message exchange validation using two different setups: the OAI-RF simulator-based \cite{rfsimulator} setup and the O-RAN-based Localization Testbed at EURECOM. The OAI-RF simulator allowed us to test the complete message flow and protocol interactions in a controlled environment, while the O-RAN-based Localization Testbed provided a real-world scenario to further validate the functionality and reliability of our UL-TDoA implementation. These setups ensure the robustness of our implementation and confirm its capability to support UL-TDoA positioning.
\subsection{OAI-rfsimulator based setup}
Our first validation approach uses a setup based on the OAI-RF simulator, a tool within the OpenAirInterface (OAI) framework designed to emulate RF environments. This virtualized environment allows for the testing of various network scenarios without physical hardware, enabling end-to-end protocol and message exchange validation. By replicating real-world RF conditions, the OAI-RF simulator supports the validation of 5G functionalities, such as UL-TDoA positioning, in a controlled and repeatable manner, making it a valuable asset for testing and refining implementations efficiently.

\begin{figure}[ht]
\centering
\includegraphics[width=\linewidth]{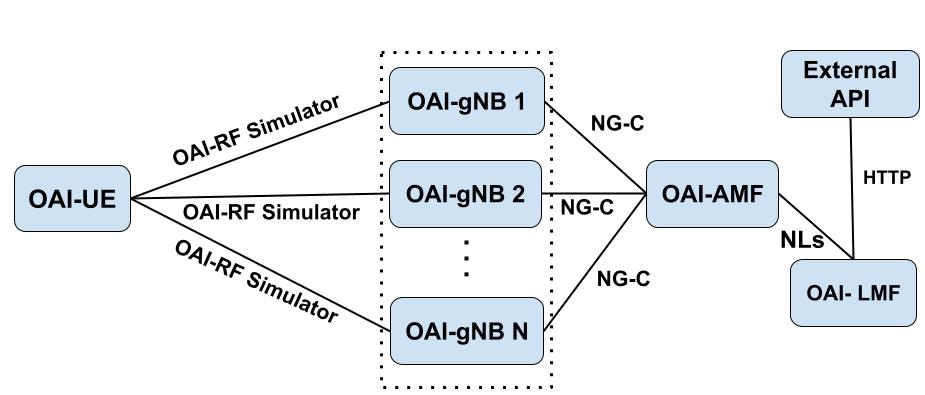}
\caption{OAI-RF simulator-based setup}
\label{fig:oairfsim}
\end{figure}

A detailed tutorial on building an OAI-RF simulator-based setup can be found here\cite{5gran_rfsimu_tutorial}. The working branch for our implementation is named \textit{NRPPA\_Procedures} \cite{5gran_nrppa}. In the following, we discuss the key aspects of our setup and provide relevant log details.

\subsubsection{Preparing the Setup}
The first step is to deploy the 5G core. In our setup, we utilized a Docker-based deployment of the OAI 5G core. The Docker images for the 5G core network functions  (NFs) are open source, publicly available, and can be pulled to the local server using the following commands.

{\scriptsize
 \begin{Verbatim}[breaklines=true, breakanywhere=true]
#Pull OAI Core Network Functions
docker pull oaisoftwarealliance/ims:latest
docker pull oaisoftwarealliance/oai-amf:develop
docker pull oaisoftwarealliance/oai-nrf:develop
docker pull oaisoftwarealliance/oai-smf:develop
docker pull oaisoftwarealliance/oai-udr:develop
docker pull oaisoftwarealliance/oai-upf:develop
docker pull oaisoftwarealliance/oai-udm:develop
docker pull oaisoftwarealliance/oai-ausf:develop
docker pull oaisoftwarealliance/oai-lmf:develop
docker pull oaisoftwarealliance/trf-gen-cn5g:latest
 \end{Verbatim}
}
The next step is to clone the OAI 5G RAN repository \cite{5gran_nrppa} and build the RFsimulator-based gNB and UE. Our work is part of the \textit{NRPPA\_Procedures} branch\footnote{The main branch of the OAI 5G RAN repository is \textit{develop}, and all new features are eventually merged into this branch. If the NRPPA\_Procedures branch no longer exists, it indicates that the features have been merged into the \textit{develop} branch of the OAI 5G RAN repository, and you should use the \textit{develop} branch instead.}.

{\scriptsize
 \begin{Verbatim}[breaklines=true, breakanywhere=true]
 git clone https://gitlab.eurecom.fr/oai/openairinterface5g.git 
 cd openairinterface5g/cmake_targets
 git checkout NRPPA_Procedures
 # install dependencies
 ./build_oai -I 
 # compile gNB and nrUE
 ./build_oai --gNB --nrUE -w SIMU 
 \end{Verbatim}
}

Once the NF images have been pulled and the RFsimulator-based gNB and UE have been successfully built, we are ready to deploy our RFsimulator-based setup using these components. Use the following commands to deploy the 5G core.
{\scriptsize
 \begin{Verbatim}[breaklines=true, breakanywhere=true]
cd openairinterface5g/doc/tutorial_resources/oai-cn5g
docker compose -f docker-compose.yaml up -d
# verify the deployment
docker ps -a
 \end{Verbatim}
}

\begin{figure}[ht]
\centering
\includegraphics[width=\linewidth]{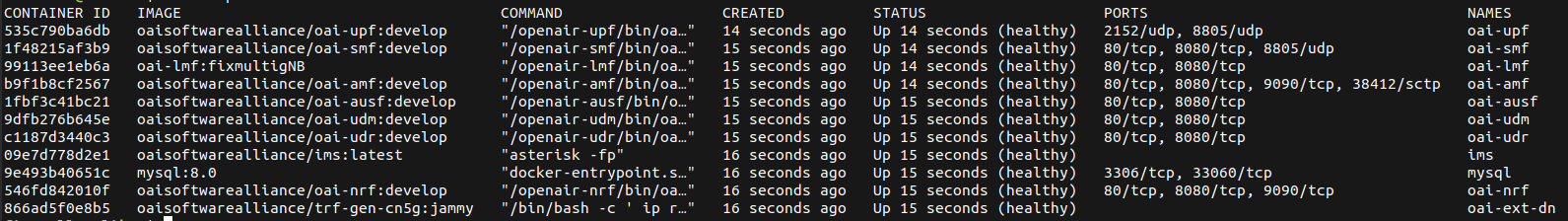}
\caption{5G core deployment logs}
\label{fig:5gcore}
\end{figure}

Figure \ref{fig:5gcore} shows the logs from a successful 5G core deployment.
The logs of the AMF can be checked using the following command.
{\scriptsize
 \begin{Verbatim}[breaklines=true, breakanywhere=true]
 docker logs oai-amf -f
 \end{Verbatim}
}

\begin{figure}[ht]
\centering
\includegraphics[width=\linewidth]{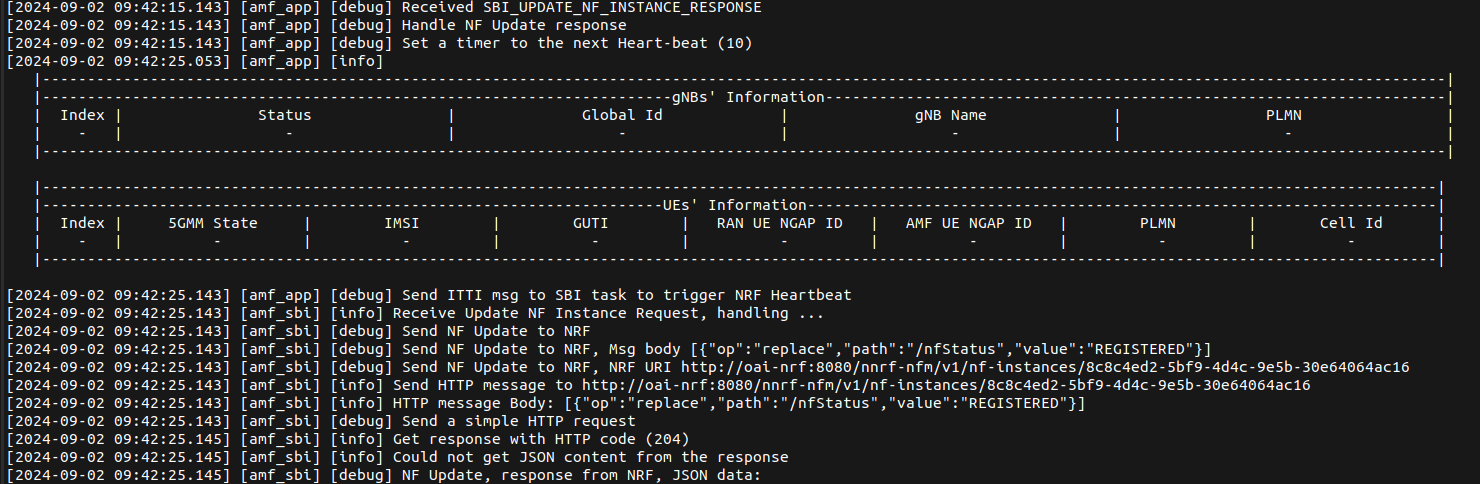}
\caption{AMF logs before running gNB and UE }
\label{fig:amf}
\end{figure}

Figure \ref{fig:amf} shows the AMF logs prior to starting the gNB and UE.

Next, we start the gNB and UE using the following commands 
{\scriptsize
 \begin{Verbatim}[breaklines=true, breakanywhere=true]
#open a terminal to run the gNB
cd openairinterface5g/cmake_targets/ran_build/build
sudo RFSIMULATOR=server ./nr-softmodem -O ../../../targets/PROJECTS/GENERIC-NR-5GC/CONF/ gnb.sa.band78.fr1.106PRB.usrpb210.conf --gNBs.[0].min_rxtxtime 6 --rfsim --sa
#open another terminal to run the UE
cd openairinterface5g/cmake_targets/ran_build/build
sudo RFSIMULATOR=127.0.0.1 ./nr-uesoftmodem -r 106 --numerology 1 --band 78 -C 3619200000 --sa --uicc0.imsi 001010000000001 --rfsim
 \end{Verbatim}
}

\begin{figure}
\centering
\includegraphics[width=\linewidth]{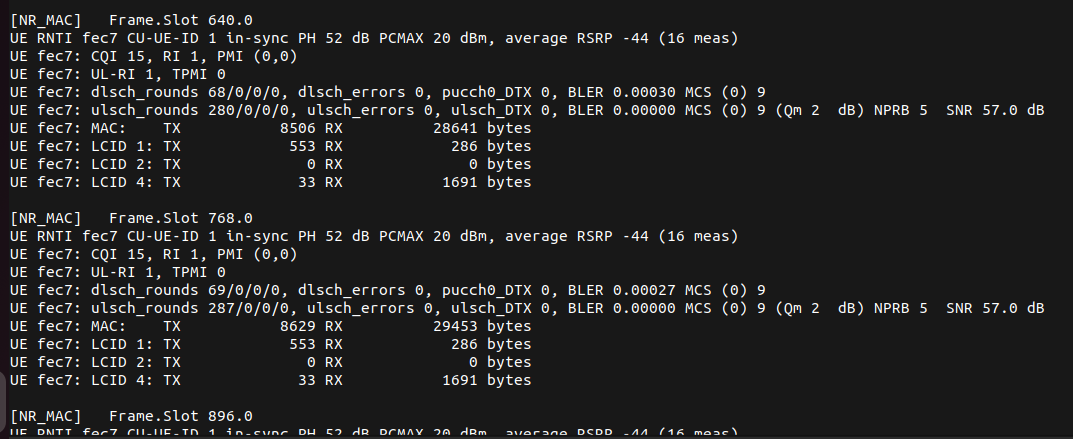}
\caption{Example logs of gNB in the UE connected state.}
\label{fig:gnbconnected}
\end{figure}
\begin{figure}
\centering
\includegraphics[width=\linewidth]{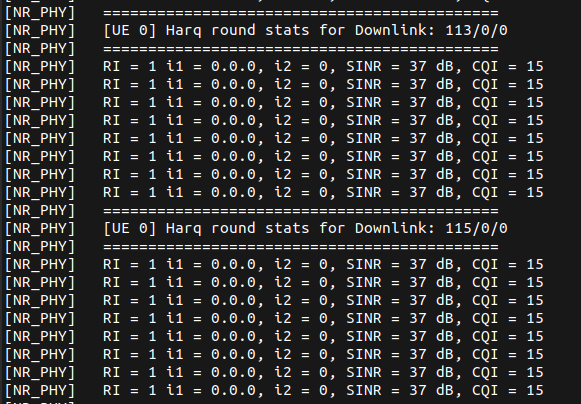}
\caption{Example logs of UE in the connected state.}
\label{fig:ueconnected}
\end{figure}
\begin{figure}
\centering
\includegraphics[width=\linewidth]{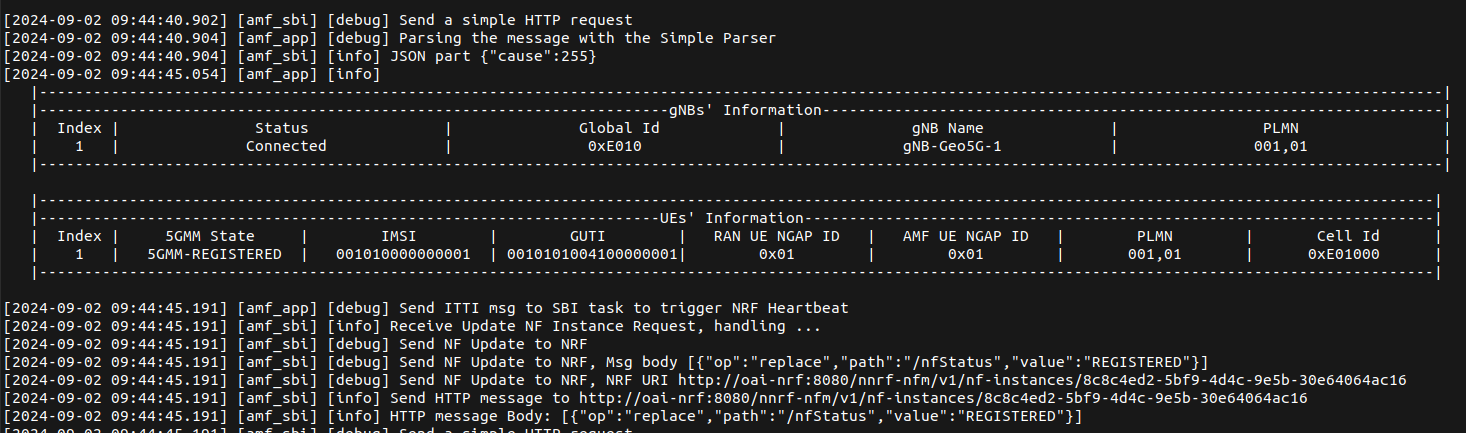}
\caption{Example logs of AMF in the UE connected state. }
\label{fig:amfconnected}
\end{figure}

Figures \ref{fig:gnbconnected}--\ref{fig:amfconnected} show example logs of the gNB, UE, and AMF when the UE is in the connected state.


Now we are ready to start the positioning procedure for the UE. To initiate the positioning process for a specific UE, we send an HTTP POST request to the determine-location API of the LMF. The request contains a data structure of type InputData \cite[Section 6.1.6.2.2]{lmf}, which includes the IMSI value of the UE. We can initiate the positioning request using the following command, where the IP address of the LMF is $192.168.70.141$.


{\scriptsize
 \begin{Verbatim}[breaklines=true, breakanywhere=true]
 curl --http2-prior-knowledge -H "Content-Type: application/json" -d "@InputData.json" -X POST http://192.168.70.141:8080/nlmf-loc/v1/determine-location
 \end{Verbatim}
}

\begin{figure}[ht]
\centering
\includegraphics[width=\linewidth]{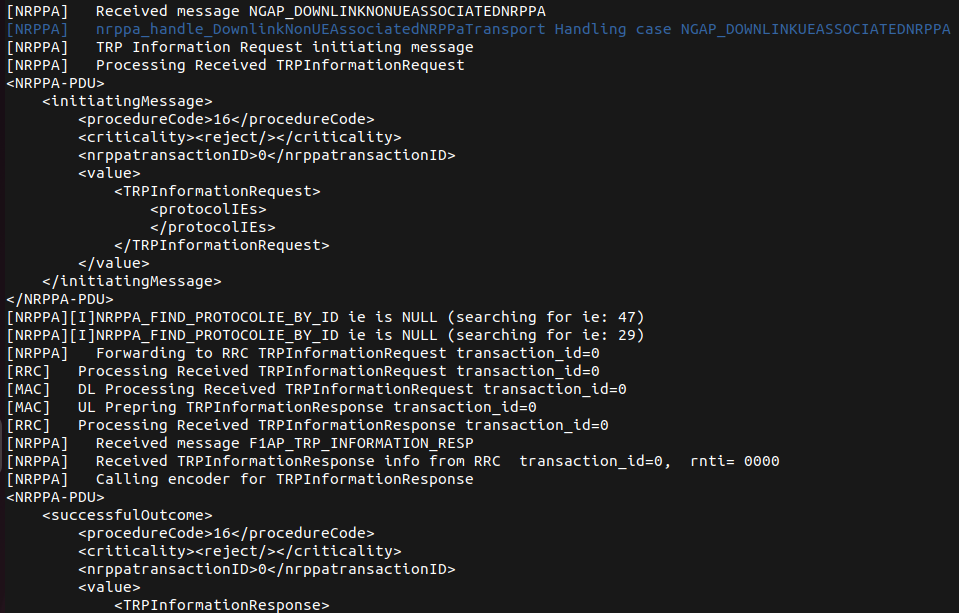}
\caption{Example logs when gNB received TRP information request }
\label{fig:gnbtrp}
\end{figure}

\begin{figure}[ht]
\centering
\includegraphics[width=\linewidth]{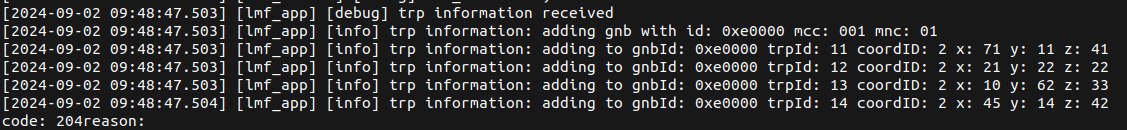}
\caption{Example logs when LMF received TRP information response }
\label{fig:lmftrp}
\end{figure}

\begin{figure}[ht]
\centering
\includegraphics[width=\linewidth]{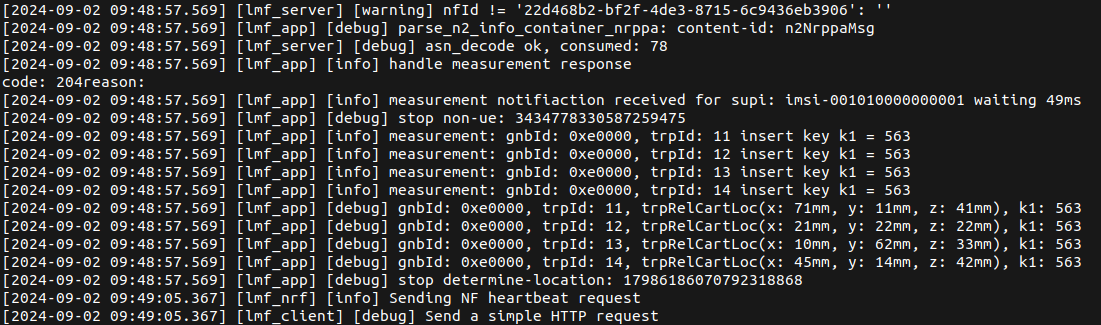}
\caption{Example logs when LMF received ToA information}
\label{fig:lmftoa}
\end{figure}

Figures \ref{fig:gnbtrp}--\ref{fig:lmftoa} show example logs of the gNB and LMF after initiating the positioning request. The LMF logs can be checked using the following command.



{\scriptsize
 \begin{Verbatim}
 docker logs oai-lmf -f
 \end{Verbatim}
}

\subsection{O-RAN based Localization Testbed at EURECOM}
The localization testbed is part of the larger Open5G testbed deployed at EURECOM. The overall testbed consists of computing and switching infrastructure interconnected through high-speed fiber links with various radios (either software defined radios like the USRP or commercial O-RAN radio units) allowing for virtualized deployment of 5G networks. A critical component of the EURECOM 5G testbed is its integration with OpenAirInterface, which provides all the virtualized network functions to run the 5G network. 

\subsubsection{Deployment}

\begin{figure}
\begin{center}
\includegraphics[width=0.9\columnwidth]{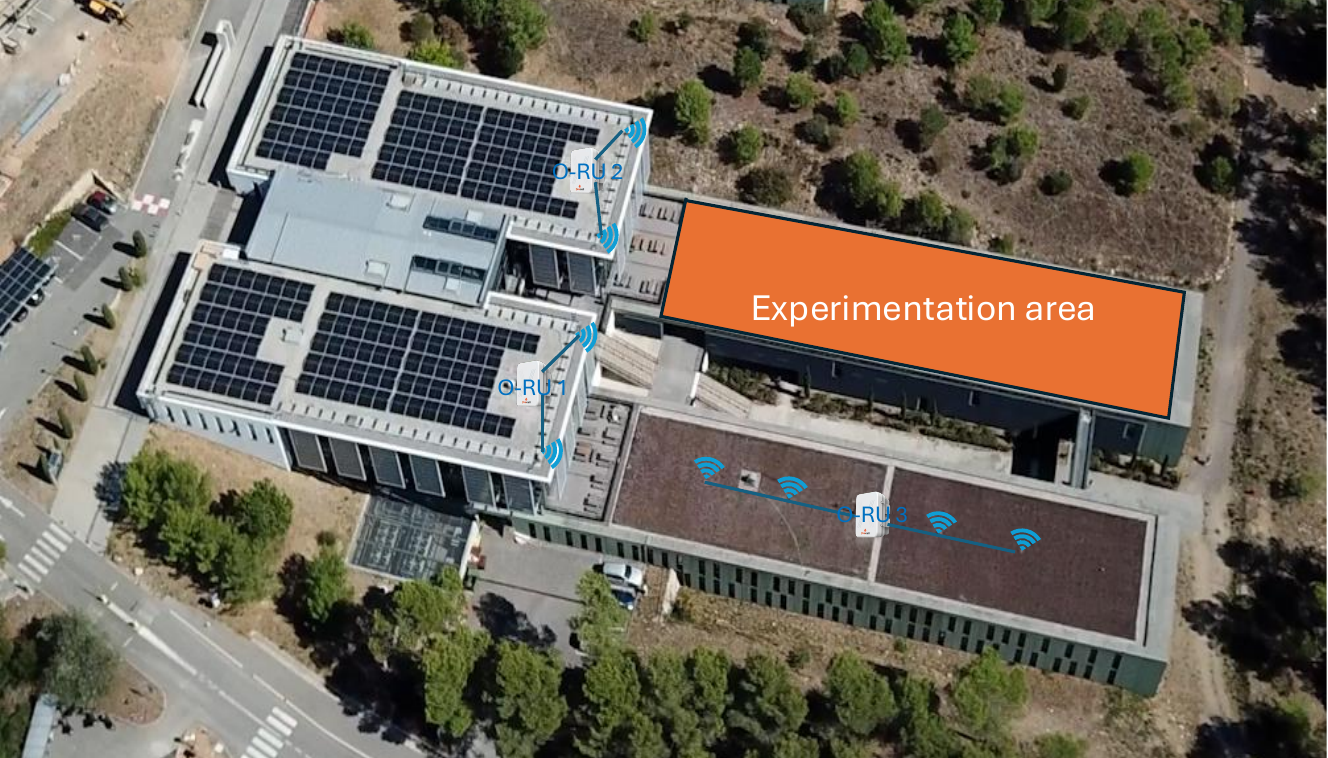}
\caption{Deployment of the localization testbed at EURECOM.}
\label{fig:deployment}
\end{center}
\end{figure}

\begin{figure}
\begin{center}
\includegraphics[width=1\columnwidth]{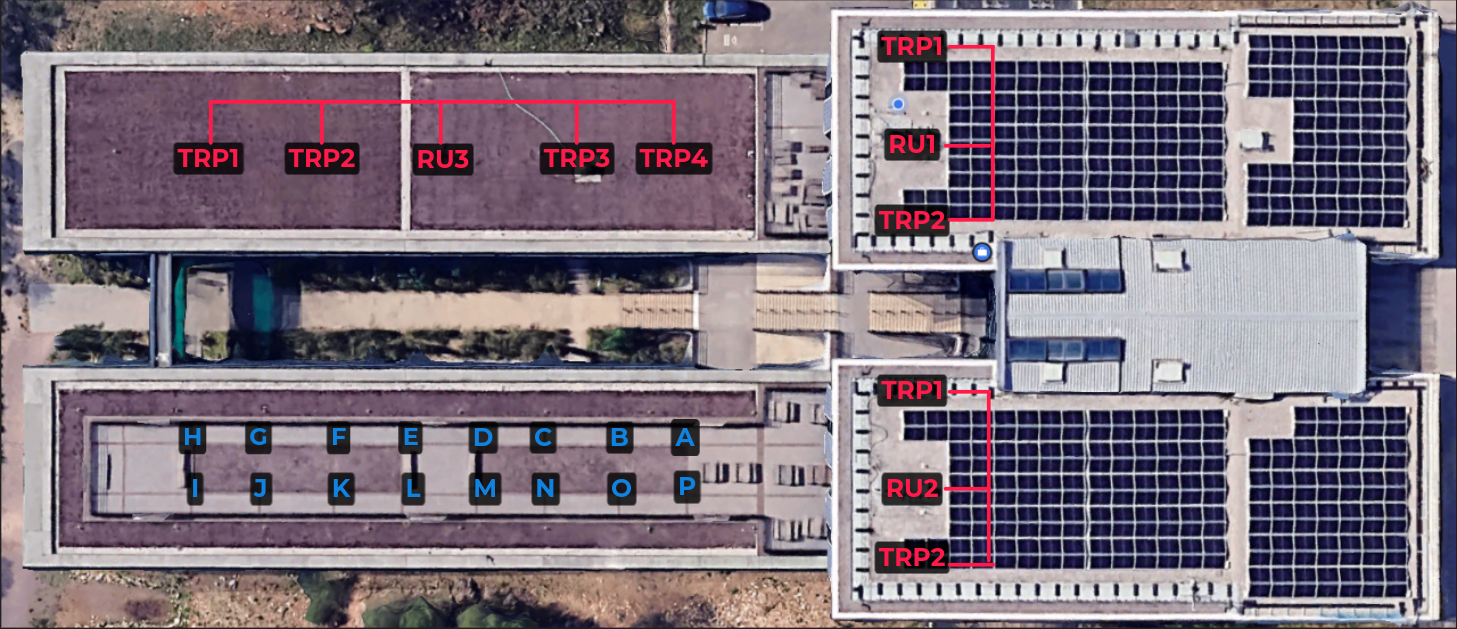}
\caption{TRPs and ground truth positions}
\label{fig:architecture}
\end{center}
\end{figure}

In order to evaluate the localization features described above, we have deployed 3 additional O-RUs from Firecell outdoors (see Table \ref{tab:ru} for details). An aerial photo of the EURECOM building is shown in Figure \ref{fig:deployment}. The building has two symmetric wings and two different levels following the slope of the hill it is built on. The top level roof features two antenna sites, one on each wing, which are part of the greater EURECOM 5G testbed. Each site has multiple fibre connections that run directly into EURECOM's server room which hosts the computing and switching infrastructure. We have deployed a Firecell RU on each of these sites and each RU drives 2 external antennas, which have been mounted on the railings of the roof and are connected to the RU through a 10m low loss cable (the cable introduces an attenuation of 7 dB, which has been included in the link budget). These antennas overlook the lower roof of the Eurecom building, which is three floors below. The northeast wing of the lower roof is accessible to people and is used as an experimentation area. The southwest wing of the lower roof is off-limits for people and features a third RU whose antennas are deployed on small tripods spaced 10m apart and at a height of 2m. Our deployment can be seen in Figures \ref{fig:deployment} and \ref{fig:architecture}.

\begin{table}
\caption{Parameters of the Firecell O-RU}
\begin{center}
\begin{tabular}{|l|l|}
\hline
\multicolumn{2}{|c|}{\bf NR Radio Specification}\\
\hline
Band & n77\\
Occupied Bandwidth(max) & 100MHz\\
Duplex Mode & TDD \\
Sub Carrier Spacing & 30KHz \\
MIMO & 4T4R \\
RF Output Power per port & 250mWatt/ 24dBm\\
Antennas & Internal/External \\
\hline
\multicolumn{2}{|c|}{\bf Connectivity Specification} \\
\hline
Physical & \makecell[l]{10G Base-T over SFP\\1G Base-T over Ethernet}\\
Interface Protocol & ORAN Split 7-2 CAT-A\\
Time and Synchronization & IEEE 1588v2, ITU T G.8275.1\\
\hline
\multicolumn{2}{|c|}{\bf Environmental Specification} \\
\hline
Powering & PoE ++ Type 3 IEEE802.3bt\\
Dimension (mm) & 250mm x 213.5mm x 92.1\\
Weight & $<$4Kg\\
Operating Temperature & -5 to 40C / -40 to 55C \\
Environmental & IP31/ IP65 \\
Mounting Style & Wall/ Pole / Ceiling \\
\hline
\end{tabular}
\end{center}
\label{tab:ru}
\end{table}%

\subsubsection{Logical Architecture}
\begin{figure}
\begin{center}
\includegraphics[width=0.9\columnwidth]{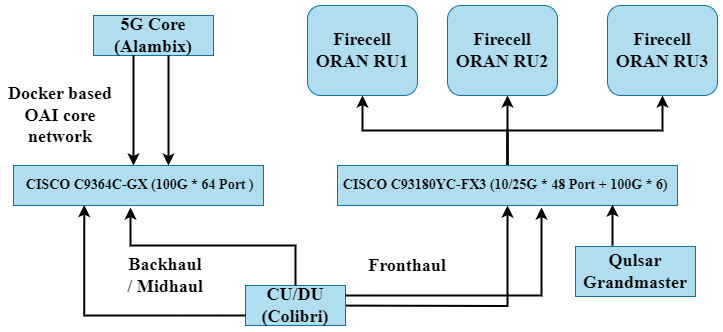}
\caption{Logical architecture of the localization testbed.}
\label{fig:logi_arch}
\end{center}
\end{figure}

The logical architecture of the localization testbed is depicted in Figure \ref{fig:logi_arch}. It follows a basic O-RAN architecture with sync option LLS-C3, i.e., the synchronization is managed by a Primary Reference Time Clock (PRTC) and a Telecom Grandmaster (T-GM) that are situated within the fronthaul network. The timing signals are distributed from the PRTC/T-GM to the O-RU and the O-DU via the fronthaul network (in our case a Cisco 93180-YC-FX3).   

The CU and DU run on a server "colibri" with a Intel(R) Xeon(R) Gold 6354 CPU @ 3.00GHz CPUs with 18 cores each. For the fronthaul network it uses an Intel X710  4x10Gbps NIC. Two of the four ports are used to connect to the backhaul network (via a CISCO C9364C-GX switch) and the two other ports are connected to the fronthaul network. The core network runs on the server "alambix" in a docker environment.

\subsubsection{Ground truth measurements}
In this work, 16 ground truth positions (A-P) on the Eurecom experimentation area were selected in a relative Cartesian coordinate system. The distances from each point to all antennas of the testbed were measured using a BOSCH GLM-150-C laser distance meter, suited for outdoor applications and long-distance measurements with a range of 0.08 to 150.00 meters. The laser was mounted on a tripod at a height of 1.3 meters, approximating the typical height when a person holds a UE. The first antenna, located on RU3 at the corner of the building, served as the reference point with coordinates (x=0, y=0, z=2.2). The positions of the other antennas were measured relative to this reference. The recorded distances were then used in nonlinear Euclidean distance equations to calculate the relative Cartesian coordinates for each ground truth point.
For all the points $P_i$ with unknown position and antennas $A_j$ with known and fixed position the following distance $d_{i,j}$ system of equations has to be solved to find $P_i = \{x_i,y_i\}$

\begin{equation}
    d_{ij} = \sqrt{(x_i - x_{j})^2 + (y_i - y_{j})^2 + (z_i - z_{j})^2}
\end{equation}
\begin{equation}
    f_i(x_i,y_i) = \sqrt{(x_i - x_{j})^2 + (y_i - y_{j})^2 + (z_i - z_{j})^2} - d_{i,j} = 0
\end{equation}

In MATLAB \texttt{fsolve} is a numerical solver used to find the roots of systems of nonlinear equations $f_i(x_i,y_i)$. When solving for the coordinates of un unknown point based on known distances to a set of antennas, \texttt{fsolve} adjusts the values of $(x_j, y_j)$ iteratively. It minimizes the difference between the calculated distances (from the guessed coordinates to the known antenna positions) and the given distances. By doing this, \texttt{fsolve} finds the point $(x_j, y_j)$ that satisfies the system of equations, effectively solving for the unknown coordinates.

\subsection{Preprocessing and Position Estimation}
In a multi-RU positioning setup, precise RU synchronization is crucial for accurate TDoA measurements, as just a 1-nanosecond error can cause a 0.3-meter positioning inaccuracy. While in some setups this accuracy can be achieved, in our GEO5G testbed we have observed timing errors of up to 40ns. 
Using a common reference TRP across all RUs will therefore result in inaccurate TDoA measurements and degraded positioning estimates. 

To mitigate this, a per-RU reference approach is adopted, where within each RU, one TRP is designated as the local reference for TDoA calculations, ensuring that all measurements within the RU remain internally consistent and unaffected by synchronization drifts in other RUs. 

Next, to ensure robust TDoA estimation while mitigating synchronization errors, a two-stage filtering mechanism is employed. Firstly, by averaging over the $N$ most recent TDoAs from each RU, we avoid small-scale fluctuations in the measurements. 
Secondly, for compensating larger spikes in the TDoAs, another filter is used that exploits the geographical information of the testing area. Knowing the minimum and maximum possible TDoA values based on known TRP positions, we define a filter that discards any invalid TDoA measurement and retains only those within a predefined bound, along with their corresponding TRP coordinates. This filtering approach effectively mitigates the impact of noisy measurements caused by multipath, Non-Line-of-Sight (NLoS) conditions, and hardware impairments. 
\subsection{Results}
\begin{figure}
\begin{center}
\includegraphics[width=1\columnwidth]{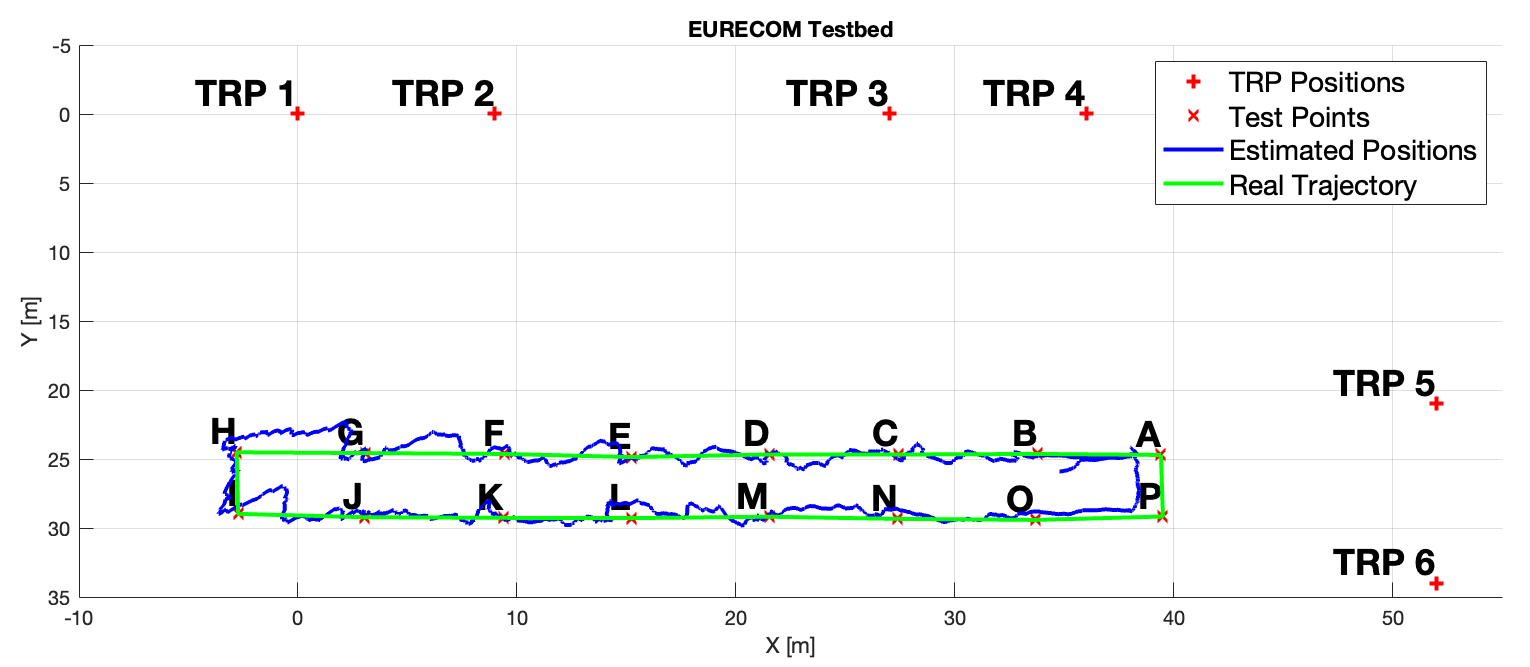}
\caption{UE tracking in a mobile scenario}
\label{fig:mobile}
\end{center}
\end{figure}

This section validates the performance of our deployed 5G network by registering a UE and triggering the location determination API. The TRP coordinates and UL-RToA values from each TRP are used in a stochastic optimization method for position estimation from our previous work \cite{DBLP:conf/ipin/AhadiMKT23}. 
The positioning error was evaluated for two setups: a single RU (RU3) and a multi-RU (RU2 and RU3). The results show that multiple RUs improve accuracy by increasing TRP diversity in both the x and y axes. In the single RU setup, the TRPs are arranged linearly, causing higher uncertainty in the y-axis, which is reduced with the multi-RU setup. Points A, B, O, and P on the edge of the testing area show degraded accuracy due to NLoS, multipath and diffraction conditions to the buildings around them. 
Table \ref{table:MAE}, is summarizing the Mean Absolute Error (MAE) for both setups. Also, Figure \ref{fig:mobile} shows a mobile scenario where a person holds the UE and takes a trajectory covering all testing points in the multi-RU setup.

\begin{table}
\centering
\resizebox{0.5\textwidth}{!}{%
\begin{tabular}{|c|c|c|c|c|c|}
\hline
\textbf{Point} & \textbf{1 RU MAE (m)} & \textbf{2 RUs MAE (m)} & \textbf{Point} & \textbf{1 RU MAE (m)} & \textbf{2 RUs MAE (m)} \\ \hline
A & 4.77 & 1.88 & I & 3.02 & 0.60 \\ \hline
B & 4.69 & 1.13 & J & 2.32 & 0.83 \\ \hline
C & 0.81 & 0.82 & K & 3.38 & 0.82 \\ \hline
D & 1.30 & 0.83 & L & 3.23 & 1.20 \\ \hline
E & 2.37 & 0.95 & M & 3.34 & 0.64 \\ \hline
F & 2.00 & 0.55 & N & 4.85 & 0.68 \\ \hline
G & 2.27 & 0.92 & O & 5.41 & 0.97 \\ \hline
H & 3.31 & 0.56 & P & 4.12 & 1.98 \\ \hline
\end{tabular}
}
\caption{MAE for single RU (RU3) and multi-RU (RU2+RU3) setups at points A-P}
\label{table:MAE}
\end{table}

In future developments, a multi-gNB setup with additional RUs and increased TRPs is expected to enhance accuracy by enabling techniques such as multipath and NLoS mitigation.

\section{Conclusions}
This paper provided the first open-source implementation of the 3GPP UL-TDoA positioning method within the OAI framework, successfully integrating UL-TDoA into both the RAN and CN components. By adhering to 3GPP standards, this implementation enabled precise and real-time positioning of UE in 5G networks, offering a flexible alternative to proprietary solutions. The approach was validated through both simulation and real-world testing, demonstrating its reliability and accuracy. This work not only enhanced the capabilities of OAI for 5G positioning but also contributed to the broader research community by providing a valuable tool for further innovation and collaboration in the field of cellular network positioning technologies.
\bibliographystyle{IEEEtran}
\bibliography{IEEEabrv,mainJ}

\ifCLASSOPTIONcaptionsoff
  \newpage
\fi

\end{document}